\documentclass{elsarticle}
\usepackage{graphics}
\usepackage{amssymb}
\usepackage{amsmath}
\usepackage{wasysym}
\usepackage{latexsym}
\usepackage{graphicx}
\usepackage{epsfig}
\usepackage{subfigure}
\usepackage{float}
\usepackage{color}
\usepackage{lineno}
\usepackage{verbatim}
\usepackage{enumitem}
\usepackage[T1]{fontenc}

\begin{document}

%\linenumbers

\begin{frontmatter}

\title{Solving the prisoner's dilemma trap in Hamilton's model of temporarily
formed random groups}

\author[mymainaddress]{Jos\'e F.  Fontanari\corref{mycorrespondingauthor}}
\cortext[mycorrespondingauthor]{Corresponding author}
\ead{fontanari@ifsc.usp.br}
  
 \author[2]{Mauro Santos}
 \ead{mauro.santos@uab.es}

\address[mymainaddress]{Instituto de F\'{\i}sica de S\~ao Carlos, Universidade de S\~ao Paulo,  13560-970 S\~ao Carlos, S\~ao Paulo, Brazil}

\address[2]{Departament de Gen\`etica i de Microbiologia, Grup de Gen\`omica, Bioinform\`atica i Biologia Evolutiva (GBBE), Universitat Aut\`onoma de Barcelona, Spain \\
cE3c - Centre for Ecology, Evolution and Environmental Changes \& CHANGE - Global Change and Sustainability Institute, Lisboa, Portugal}

\begin{abstract}
Explaining the evolution of cooperation in the strong altruism scenario, where a cooperator does not benefit from her contribution to the public goods, is a challenging problem that requires positive assortment among cooperators  (i.e., cooperators must tend to associate with other cooperators)  or punishment of defectors.  The need for these drastic measures stems from the analysis of a group selection model of temporarily formed random groups introduced by Hamilton nearly fifty years ago to describe the fate of altruistic behavior in a population.  Challenging conventional wisdom, we show analytically here that strong altruism evolves in Hamilton's original model in the case of biparental sexual reproduction. Moreover, when the cost of cooperation is small and the amplified contribution shared by group members is large, cooperation is the only stable strategy in equilibrium.  Thus, our results provide a solution to the  `problem of origination'  of strong altruism, i.e. how cooperation can take off from an initial low frequency of cooperators.
 We discuss a possible reassessment of cooperation in cases of viral co-infection, as cooperation may even be favored in situations where the  prisoner's dilemma applies.
\end{abstract}

\end{frontmatter}

\section{Introduction}

As  noted by Darwin in his discussion of courage and self-sacrifice, traits associated with altruistic behavior are counter-selected within a group, but increase the group's productivity and the group's chances of survival in the case of intergroup fighting \cite{Darwin_1871}. Recognizing that the possible outcomes of this conflict between different levels of selection could only be determined by mathematical analysis, two slightly different models of group selection were proposed nearly fifty years ago to study the conditions for the evolution of altruistic behavior \cite{Hamilton_1975,Wilson_1975}.  Both models assume temporary random grouping of individuals, but differ in the way in which within-group competition between selfish and altruistic individuals is implemented. Hamilton's model follows the Darwinian view  that altruists are always at disadvantage in local competition with selfish individuals, but groups with more altruists contribute more offspring to a pool of dispersers that will eventually form the new groups \cite{Hamilton_1975}. Wilson's trait group model ignores within-group competition altogether: although it assumes that individuals acquire their fitness locally, the competition between altruistic and selfish individuals takes place globally, i.e., in the population at large \cite{Wilson_1975,Wilson_1990}.  Recently, we have pointed out the equivalence between Wilson's model and the evolutionary game approach to studying the emergence of cooperation in $n$-person public goods games, which is based on the replicator equation framework \cite{Fontanari_2024a}.   We are aware of the distinction between cooperation, which involves mutual benefit, and altruism, which implies `others-only' benefit \cite{Pepper_2000, West_2007}. However, we will use both terms, cooperation and altruism, more or less interchangeably here.

The outcome of these group selection  models depends on whether an altruist benefits from her altruistic behavior or not.   When the return from altruistic behavior exceeds the cost of performing it, we have the scenario of weak altruism  \cite{Wilson_1990,Okasha_2009}. This is still altruism in the sense that  selfish individuals benefit the most from this behavior.  The traditional $n$-person prisoner's dilemma \cite{Fox_1978,Boyd_1992,Fehr_2000} and the $n$-person snowdrift game \cite{Fontanari_2024a,Zheng_2007,Santos_2012,Archetti_2012} fall into this category. Both Hamilton's and Wilson's models of group selection with randomly formed groups predict that altruism will evolve in this scenario, a clearly impossible outcome for homogeneous or unstructured populations,  which explains the notoriety of  these models. When an  altruist has no return from her altruistic behavior, as in the others-only variant of the $n$-person prisoner's dilemma \cite{Hamilton_1975,Hannelore_2010,Sigmund_2010}, we have the more challenging scenario of  strong altruism.  Hamilton's seminal paper addressed this scenario and concluded that altruism cannot progress in randomly formed groups: cooperation is impossible unless there is some form of positive assortment among cooperators \cite{Hamilton_1975}.  Wilson came to the same conclusion when groups are formed by binomially sampling individuals from the total population, as in Hamilton's model, but mistakenly suggested that environmental heterogeneity could produce a sufficiently large excess of variance over the binomial expectation that strong altruism might be favored \cite{Wilson_1975}. This error was quickly corrected \cite{Charlesworth_1979}, so that both models now point to the impossibility of strong altruism evolving in the absence of explicit positive assortment among altruists, i.e., altruists must have a tendency to group  with other altruists.  Nonetheless, we show here, using both analytical and simulation results, that strong altruism can evolve in Hamilton's model  without assuming positive assortment. This is achieved by not taking  the phenotypic gambit in full \cite{Grafen_1984}.

The phenotypic gambit, often used in evolutionary game theory, assumes, among other things, the simplest genetic system (e.g., haploid individuals). Although, this  is usually wrong, we can generally ignore genetics to study the evolutionary fate of behavioral strategies, except in the case of diploidy, when the heterozygote is the fittest genotype \cite{Grafen_1984} (see also \cite{Broom_2013}). Let us  accept the gambit and assume that organisms are haploid and that each chromosome consists of $L$ loci, one of which controls one of the only two pure strategies, altruistic (cooperator) and selfish (defector). Now consider three possible modes of reproduction that are commonly used in theoretical models: asexual reproduction, sexual reproduction (production of haploid gametes that combine in pairs to produce a recombinant offspring chromosome) that allows self-mating, and sexual reproduction that requires two different individuals, what Williams  calls euphrasic \cite{Williams_1975}. 
Thus, mating pairs are formed by randomly drawing individuals (with or without replacement) according to their relative fitness, and each pair produces a single offspring.  This is still a simplified picture of a more complex reality, where organisms can be diploid and the behavioral strategy is likely to be controlled by many loci with possible epistatic interactions. The question here is whether the phenotypic gambit, which in this case would also ignore the complications introduced by the mode of reproduction, is still valid. If so, we would conclude that strong altruism cannot evolve in Hamilton's model regardless of how individuals reproduce. 

Here, we show that the way we model reproduction can have a qualitative impact on the evolution of altruism in ways that the phenotypic approach cannot predict or explain. In particular, we show that strong altruism can evolve under biparental sexual reproduction in Hamilton's model \cite{Hamilton_1975}. Strictly speaking, and considering eukaryotes, we analyzed the fate of altruism in a monoecious population, where individuals have both male and female reproductive organs and can produce gametes of both sexes, but cannot self-fertilize.  Earthworms, for example, are monoecious.  This nondifferentiation between the sexes simplifies the model and allows a full analytical solution in the infinite population limit. In an appendix, we use simulations to extend the analyses to a dioecious population (with distinct unisexual individuals) and show that sex differentiation actually promotes altruistic traits.

The rest of the paper is organized as follows.  In Section \ref{sec:mod} we  describe Hamilton's model of temporarily
formed random groups,  assuming $G$ groups of size $n$, using the terminology of the $n$-person prisoner's dilemma.  In Section \ref{sec:det} we develop an analytical treatment to solve Hamilton's model in the deterministic limit $G \to \infty$ and show that in the case of asexual reproduction, cooperation is doomed to extinction in the strong altruism scenario, but not in the weak altruism scenario. In addition,  we show that sexual reproduction, allowing self-mating, leads to the same conclusion as in the asexual case. We then consider biparental sexual reproduction and, challenging conventional wisdom, derive the conditions that guarantee the emergence of cooperation in the strong altruism scenario. In all cases, we follow the fate of the altruistic or selfish strategy, which for sexual reproduction would be analogous to following the allelic changes in the recombinant offspring chromosomes hosting the behavioral gene.  In Section \ref{sec:fin} we use simulations to prove the robustness of the conclusions of the deterministic analysis to the effect of demographic noise that arises when $G$ is finite. 
Finally,  in Section \ref{sec:disc}  we summarize our results and also suggest a possible reassessment of cooperation in cases of viral co-infection, as cooperation may even be favored in situations where the  prisoner's dilemma applies. In \ref{ref:A} we present the simulations for a dioecious population with distinct unisexual individuals.  Returning to the phenotypic gambit, we cannot avoid ending this section by recalling a quote from Albert Einstein (possibly misattributed; see \cite{Robinson_2018}): ``Make things as simple as possible, but not simpler.''.

\section{Model}\label{sec:mod}

Here we recast Hamilton's model of temporarily formed random groups \cite{Hamilton_1975} using the terminology of  public goods games \cite{Fox_1978}.
Consider a population of $Gn$ individuals, divided into $G$ groups, each of which contains $n$ individuals. Within each group, individuals play a $n$-person prisoner's dilemma game to determine their fitness: each individual may (or may not) contribute an amount $c>0$ to the public goods, which is then multiplied by a factor $r>1$, and the resulting amount $r c$ is divided among the $n-1$ other players. This is the others-only variant of the $n$-person prisoner's dilemma, where contributors do not benefit from their own contributions \cite{Hannelore_2010}.   We refer to individuals that contribute to the public goods as cooperators and those that do not as defectors. 

The fitness (or payoff) of a cooperator   in a group with $i-1$ other cooperators is
\begin{equation}\label{fa}
f_c (i) = 1 - c + (i-1) \frac{rc}{n-1} ,
\end{equation}
whereas the fitness  of a defector  in a group with $i$ cooperators is
\begin{equation}\label{fs}
f_d (i) = 1 + i \frac{rc}{n-1} .
\end{equation}
These are the payoffs of the popular $n$-person prisoner's dilemma \cite{Fox_1978,Hannelore_2010,Sigmund_2010} with a constant baseline fitness value of $1$. They  are identical to the fitness attributed to altruistic and selfish individuals in Hamilton's model of randomly formed groups: in  Hamilton's terminology, an altruist gives up $k=c$ units of her own fitness in order to add $K=rc$ units  to the joint fitness of her $n-1$ companions \cite{Hamilton_1975}. We will assume $c \in [0,1]$ to ensure that the fitness values are not negative, so   $c=1$ is the worst scenario for cooperation.  
The parallel between the  $n$-person prisoner's dilemma  and Hamilton's model ends here, however.  In fact,  in Hamilton's model the fitness of  a group with $i$  cooperators, 
\begin{equation}\label{pi}
\pi(i) = n + i c (r-1) ,
\end{equation}
determines the number of  offspring $m$ produced by the group. In particular, we assume that $m$ is a random variable distributed by a Poisson distribution with mean $\pi(i)$.    Group fitness plays no role in the $n$-person prisoner's dilemma, where individuals decide whether or not to keep their strategies by comparing their payoffs to those of randomly selected individuals in the population at large \cite{Traulsen_2005,Fontanari_2024b}, as in Wilson's trait group formulation \cite{Wilson_1975,Wilson_1990}, so there is effectively no within-group competition. 

In Hamilton's model, within-group competition takes place during the reproduction stage. As mentioned before, we will consider three modes of reproduction. In the asexual reproduction mode, which is the one considered in Hamilton's seminal paper, we select one parent within each group with probability equal to her relative fitness, and the single offspring inherits the strategy of the parent.  For each group, exactly $m$ parents are selected in this way, and the resulting $m$ offspring are added to the offspring pool. Hamilton refers to the offspring pool as the migrant pool, which is formed by young individuals that reach maturity \cite{Hamilton_1975}.  In sexual reproduction with selfing allowed, we randomly select two not necessarily different parents with probability given by their relative fitness, and the single offspring inherits the strategy of either parent with equal probability. This selection process is repeated $m$ times resulting in the addition of $m$ offspring to the offspring pool. This mode is clearly equivalent to asexual reproduction,  as we will show in the following section. Finally, in the biparental sexual reproduction mode, the two parents, randomly chosen with probability given by their relative fitness, must be different, and the single offspring, as before, inherits the strategy of either parent with equal probability.   In this mode,  it is guaranteed that  in a group with $i=n-1$ cooperators, one cooperator will be chosen as a parent. Note that the groups contribute different numbers of offspring to the offspring pool, not only because $m$ is a Poisson distributed random variable, but also because the parameter of the Poisson, i.e. $\pi(i)$, depends on the number of cooperators within the group, which varies from group to group. 
Finally, each group of the next generation is formed by sampling $n$ individuals with replacement from the offspring pool, and the whole process is repeated until one of the strategies disappears and evolution stops.

In the others-only variant of the $n$-person prisoner's dilemma, altruistic behavior (i.e., contribution to  public goods) reduces the fitness of the cooperator, corresponding to the strong altruism  scenario. As noted above, the conventional wisdom in this case is that cooperation cannot evolve in randomly formed groups that last less than one generation \cite{Fletcher_2004}.  The more traditional variant of the $n$-person prisoner's dilemma allows the amplified amount $rc$ to be redistributed to all players, so that the cooperator gets the return $cr/n$ \cite{Boyd_1992}. If $cr/n - c> 0$ or $r >n$, then altruistic behavior confers a net benefit to the cooperator, corresponding to the weak altruism scenario. A similar situation occurs in the $n$-person snowdrift game when the payoff for completing a task is greater than the effort expended \cite{Fontanari_2024a}.  In this case, the cooperator may have a net fitness advantage over the population as a whole, despite being at a disadvantage relative to defectors within the same group.    This is the only  situation where Wilson's trait group formulation \cite{Wilson_1975,Wilson_1990} or its game theoretic counterparts \cite{Boyd_1992} can explain the maintenance of cooperation \cite{Charlesworth_1979}.

 %In the following, we study this model analytically in the deterministic limit $G \to \infty$ and by simulations in the case of finite number of groups.

 %
 \section{Deterministic limit}\label{sec:det}

 In his seminal paper \cite{Hamilton_1975}, Hamilton studied the asexual reproduction mode in the case there are infinitely  many groups  using  Price's equation \cite{Price_1970}. Here we offer an elementary derivation of  a recurrence for the frequency of cooperators $q_t$ in the offspring pool in generation $t$, which can be easily generalized to more complex scenarios and reproduction modes \cite{Alves_2000}.  
 
Given $q_t$, the proportion of groups  with $i$ cooperators among the infinitely many groups is given by the binomial distribution
\begin{equation}\label{Y}
Y_i(t) = \binom{n}{i} q_t^i (1-q_t)^{n-i} ,
\end{equation}
and our problem is to determine how many cooperators, on average, a group with $i$ cooperators contributes to the offspring pool.  We recall that, according to Section \ref{sec:mod}, a group with $i$ cooperators contributes $m$ offspring to the offspring pool with $m$  distributed by the Poisson distribution 
 \begin{equation}\label{Poisson}
p[m; \pi(i)] =   \exp [ -\pi(i)] ~\frac{[\pi(i)]^m}{m!} .
 \end{equation}
Let $Q_c(i)$ denote the probability that an offspring is a cooperator. Calculating this probability requires  to be explicit about the mode of reproduction, and we will do this calculation separately for each mode in the following, but for now let  us assume that $Q_c(i)$ is known.  The probability that there are $j$ cooperators among the $m$ offspring produced by  a group with $i$ cooperators  is given by the binomial distribution
 \begin{equation}
B[m, Q_c(i)] = \binom{m}{j} \left [Q_c(i)\right ]^j \left [1 - Q_c(i)  \right ]^{m-j}.
\end{equation}
Therefore, a  group with $i$ cooperators contributes, on average,  with
 \begin{equation}\label{av_of}
  \sum_{m=0}^\infty p[m; \pi(i)]  ~\sum_{j=0}^m B[m, Q_c(i)]  j =  \sum_{m=0}^\infty p[m; \pi(i)]  m Q_c(i) = \pi(i) Q_c(i) 
\end{equation}
cooperators to the offspring pool. Note that only the mean of the distribution of the number of offspring $m$ goes into this calculation, so any  distribution with mean $\pi(i)$ gives the same result.

The proportion of cooperators in the offspring pool in generation $t+1$ is given by the ratio between the mean number of cooperator offspring produced by the groups formed by sampling the offspring pool in generation $t$ and the mean total number of  offspring produced by these groups, i.e.
\begin{equation}\label{rec_g}
    q_{t+1} = \frac{\sum_{i=0}^n Y_i(t) \pi(i) Q_c(i)}{\sum_{i=0}^n Y_i(t) \pi(i)}.
\end{equation}
     The sum in the denominator can be easily performed and yields $ n [ 1 + c(r-1) q_t ]$. 
The last step to obtain the recurrence for $q_t$ is to calculate $Q_c(i)$ for the different reproduction modes.
 
 \subsection{Asexual reproduction}\label{sec:asex}
 
 In this case,  we randomly select a parent within a group with $i$ cooperators according  to her relative fitness, and the single offspring inherits her strategy. Thus, the probability of an offspring being a cooperator is simply
 \begin{equation}
Q_c(i) = \frac{ i f_c(i)}{\pi(i)}.
\end{equation}
Inserting this probability into Eq,\ (\ref{rec_g}) yields
  \begin{equation}\label{q_as}
 q_{t+1} =  q_t  \frac{ 1 -c +rcq_t}{1 + cq_t (r-1) } ,
 \end{equation}
 which is consistent with the recurrence derived using Price's equation, provided that the mean fitness in \cite{Hamilton_1975} is corrected to $w  = 1 +q_t(K-k) = 1 + q_t c(r-1)$. The only fixed points of  recurrence (\ref{q_as})  are $q^*=0$ and $q^*=1$. For $q_t \ll 1$ it is rewritten as $q_{t+1}  \approx (1-c) q_t$, which implies that $q_{t+1} < q_t$ and so $q^*=0$ is always stable and $q^*=1$ is always unstable.
  This conclusion agrees with  Hamilton's that cooperation cannot progress in such a model  \cite{Hamilton_1975}. 
  
 It is instructive to briefly discuss  the results for the weak altruism version of the $n$-prisoner's dilemma. In this case, the fitness of a cooperator is $\hat{f}_c (i) = 1 - c + i rc/n$ and the fitness of a defector is $\hat{f}_d (i) = 1 + i rc/n$, where $i$ is the number of cooperators in the group of size $n$. The fitness of this group is still given by Eq.\ (\ref{pi}). Thus, the recurrence for the frequency of cooperators $\hat{q}_{t} $ in the  weak altruism scenario is
   \begin{equation}\label{q_as2}
\hat{q}_{t+1} =  \hat{q}_t  \frac{ 1 -c + rc/n + (n-1) rc \hat{q}_t/n}{1 + c\hat{q}_t (r-1)} ,
 \end{equation}
 whose only  fixed points are $\hat{q}^* = 0$ and $\hat{q}^* = 1$. For $\hat{q}_{t}  \ll 1 $ we have $\hat{q}_{t+1}  \approx (1-c + rc/n ) \hat{q}_t$ so $\hat{q}^* = 0$ is unstable (and thus $\hat{q}^* = 1$ is stable) for $r > n$. This means that cooperators at low frequency can invade a resident population of defectors if the amplification factor $r$ is sufficiently large.

   \subsection{Sexual reproduction with self-mating}
   
 In this case, we randomly select two parents with replacement according to their relative fitness, and the single offspring inherits the strategy of either parent with equal probability. Since there are no restrictions on the choice of parents, the probability that an offspring is a cooperator is 
 \begin{eqnarray}
Q_c(i)  & = & \left ( \frac{ i f_c(i)}{\pi(i)} \right )^2  + \frac{1}{2}   \frac{ i f_c(i)}{\pi(i)}  \frac{ (n-i) f_d (i)}{\pi(i)}  + \frac{1}{2}   \frac{ (n-i) f_d (i)}{\pi(i)}  \frac{ i f_c(i)}{\pi(i)}  \nonumber \\
& = &  \frac{ i f_c(i)}{\pi(i)} ,
\end{eqnarray}
which is identical to the result for asexual reproduction, as expected. Thus, cooperation cannot progress also in the model where sexual reproduction with selfing is allowed.

 \subsection{Biparental sexual  reproduction}
 
 The difference to the previous case is that the mating parents within the group of size $n$ must be different individuals. Their single offspring inherits the strategy of either parent with equal probability. So the problem we have to solve now is to find the probability that the offspring will be a cooperator when the two parents are chosen from a group with  $i$ cooperators and $n-i$ defectors. There are three possibilities we need to consider, which are described below.
 
 \begin{enumerate}[label=(\alph*)]
 %--------------------------------------------
  \item Both parents are cooperators The probability that the first selected parent is a cooperator  is simply
  \begin{equation}
  \frac{if_c(i)} {if_c(i) + (n-i)f_d(i)}.
  \end{equation}
 Given that the first selected parent is a cooperator, the probability that the second is also a cooperator is 
  \begin{equation}
  \frac{(i-1)f_c(i)} {(i-1) f_c(i) + (n-i)f_d(i)},
  \end{equation}
  since the fitness $f_c(i)$ and $f_d(i)$ given by Eqs.\ (\ref{fa}) and (\ref{fs})  are computed before the  selection procedure. Thus, the probability that the two  selected parents are cooperators is 
  \begin{equation}
   \frac{(i-1)f_c(i)} {(i-1) f_c(i) + (n-i)f_d(i)} \times \frac{if_c(i)} {if_c(i) + (n-i)f_d(i)}.
  \end{equation}
  Of course, the offspring resulting from this mating is a cooperator with probability one. The probabilities of the other events are calculated using the same reasoning.
    %--------------------------------------------
  \item The first chosen parent is a cooperator and the second is a defector. The probability of this event is
  \begin{equation}
    \frac{(n-i)f_d(i)} {(i-1) f_c(i) + (n-i)f_d(i)} \times \frac{if_c(i)} {if_c(i) + (n-i)f_d(i)}
   \end{equation}
  and the resulting offspring is a cooperator with probability $1/2$.
      %--------------------------------------------
   \item The first chosen parent is  a defector and the second is a cooperator. The probability of this event is
  \begin{equation}
    \frac{i f_c(i)} {i f_c(i) + (n-i-1)f_d(i)} \times \frac{(n-i) f_d(i)} {if_c(i) + (n-i)f_d(i)}
   \end{equation}
   and the resulting offspring is a cooperator with probability $1/2$. 
 \end{enumerate}
  Combining these probabilities, we obtain that the probability the offspring is a cooperator when  two parents are chosen from a group with $i$  cooperators and $n-i$ defectors is 
  \begin{eqnarray}\label{Qa}
  Q_c(i)  &=  & \frac{i(i-1) [f_c(i)]^2} {\left [ (i-1) f_c(i) + (n-i)f_d(i) \right ] \left [ if_c(i) + (n-i)f_d(i) \right ]}  \nonumber \\
  &  &  +\frac{1}{2}  \frac{i (n-i)f_c(i)f_d(i)} {\left [ (i-1) f_c(i) + (n-i)f_d(i) \right ] \left [ if_c(i) + (n-i)f_d(i) \right ]}  \nonumber \\
  & &  +\frac{1}{2}  \frac{i (n-i)f_c(i)f_d(i)} { \left [ i f_c(i) + (n-i-1)f_d(i) \right ] \left [ if_c(i) + (n-i)f_d(i) \right ]} .
   \end{eqnarray}
    Clearly, $Q_c(0) = 0$ and $Q_c(n)=1$.  Inserting $Q_c(i) $ into the recurrence (\ref{rec_g})  yields
  \begin{equation}\label{map}
  q_{t+1} = F_n (q_t)
  \end{equation}
  with 
  \begin{equation}\label{Fn}
F_n (q_t) = \frac{1}{ n [ 1 + c(r-1) q_t ]} \sum_{i=0}^n \binom{n}{i} q_t^i (1-q_t)^{n-i} \pi(i) Q_c(i),
  \end{equation}
  where the dependence on $q_t$ is made explicit. Since $Q_c(0)=0$ we have $F_n(0) = 0$ and so $q^* =0$ is a fixed point of the recurrence (\ref{map}). In addition, noting that $Q_c(n)=1$ and $\pi(n) = n[1+c(r-1)]$ we obtain $F_n(1) = 1$, which implies that $q^* =1$ is also a fixed point of the recurrence (\ref{map}).   Whenever these two extreme fixed points are stable, there must also be an unstable coexistence fixed point.  If both extreme fixed points are unstable, then the coexistence fixed point is stable, but this situation never happens, as we will see below.  We recall that the condition for the local stability of a fixed point is $|F_n'(q^*)| < 1$, where the prime indicates the derivative of $F_n(q)$, as usual \cite{Britton_2003}.
  
 Before considering the general case of group size $n$, it is instructive to write the recurrence (\ref{map}) for $n=2$ explicitly. We have 
  \begin{equation}
  q_{t+1} = \frac{q_t}{1+c(r-1)q_t} \left [ 1 +\frac{c(r-1)}{2}(1+q_t)    \right ],
    \end{equation}
  so that the only fixed points are $q^*=0$ and $q^*=1$. Noting that  
  \begin{equation}
  F_2'(0)  =  1 +\frac{c(r-1)}{2} > 1
    \end{equation}
  and 
  \begin{equation}
F_2'(1)  = 1 -\frac{1}{2} \frac{c(r-1)}{[1+c(r-1)]}  < 1,
    \end{equation}
 we conclude that the  all-defectors fixed point is unstable and the all-cooperators fixed point is stable if $r>1$. This is expected, since the advantage of the selfish strategy comes only from within-group competition, which for $n=2$ only occurs in groups with one cooperator and one defector. However, the requirement that mating involves both individuals gives $Q_c(1)=1/2$, so this advantage is lost for $n=2$. Now, if we consider that a group formed by two cooperators contributes on average $2 + 2 c(r-1)$ offspring to the offspring pool, while a group formed by two defectors contributes on average only $2$ offspring, the dominance of the cooperators is not surprising (see also \cite{Williams_1957}). However, for groups of size $n > 2$ the situation is not so clear-cut, as we will discuss in the following.

  Let us first consider the all-defectors fixed point, $q^* =0$, for general $n$.  A trite calculation yields $F_n'(0) = \pi(1) Q_c(1) > 0$, which is rewritten as 
  \begin{equation}
  F_n'(0) =  \frac{1}{2} (1-c) \left [ 1 + \frac{ n-1 +rc}{  n- 1 +rc - c - rc/(n-1)} \right ] ,
    \end{equation}
  where we have used Eqs. (\ref{pi}) and (\ref{Qa}) with $i=1$. The condition $ F_n'(0) < 1$ is satisfied when 
  \begin{equation}\label{c_stab_s}
  c > \frac{1}{2n-3}
     \end{equation}
  regardless of the value of the amplification factor $r$. However, if this inequality is violated, then the condition $F_n'(0) < 1$ is satisfied, provided that $r< r_s$ where
  \begin{equation}\label{r_stab_s}
 r _s   = (n-1) \frac{2n -3-c}{ 1 - (2n - 3) c}.
     \end{equation}
  For small $c$, this equation  reduces to $r_s = (n-1)(2n-3) + 4(n-2)(n-1)^2 c$. 
 The important result from the `problem of origination'  perspective \cite{Wilson_1997} is that the all-defectors fixed point is unstable if inequalities (\ref{c_stab_s}) and $r < r_s$  are violated simultaneously. For example, for $n=4$ the all-defectors fixed point is stable for $c > 0.2$, regardless the value of $r$, but  for $c=0.1$  it is unstable if  $r >29.4$. It is clear from these inequalities that increasing the group size $n$ favors the selfish strategy. 
  
  The analysis of the all-cooperators fixed point, $q^*=1$, is more laborious. In this case we have
  \begin{equation}\label{Fp1g}
  F_n'(1) = n - \frac{c(r-1)}{1 + c(r-1)}  
- \frac{  \pi(n-1)  Q_c(n-1)}{ 1 + c(r-1) },
  \end{equation}
  which can be easily evaluated numerically using Eqs. (\ref{pi}) and (\ref{Qa}) to determine the regions in the parameter space where  the all-cooperators fixed point
 is stable, i.e., where the condition $|F_n'(1) | < 1$ is satisfied.  However, in the limit $c \to 0$  we can write down an explicit expression for $ F_n'(1)$, viz.,
  \begin{equation} 
  F_n'(1) = 1 + c \frac{2n-3}{2(n-1)} -  rc \frac{1}{2(n-1)^2} 
  \end{equation}
 from which it follows that the all-cooperators  fixed point  is stable for $r > r_a$ where $r_a = (n-1)(2n-3)$ in this limit.  Note that $r_a \to r_s$  for $c \to 0$.  Although for $c=0$ the recurrence (\ref{map}) reduces to $q_{t+1} = q_t$ so that the dynamics freezes at the initial conditions, a vanishingly small value of  $c$ is sufficient to lead the dynamics to $q^*=1$ if $r > r_a$ or to $q^*=0$ if $r < r_s$.

  %-----------------------------------------------------
\begin{figure}[th] 
\centering
 \includegraphics[width=.8\columnwidth]{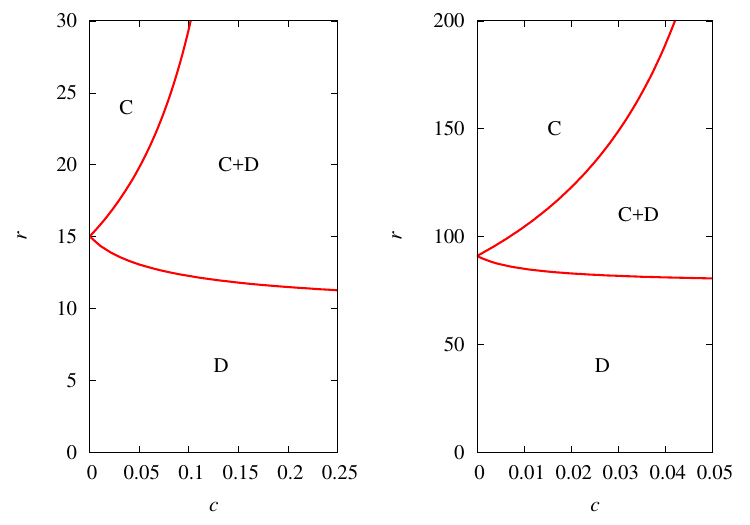}  
\caption{Phase diagram for $n=4$ (left panel) and $n=8$ (right panel)  showing the stability regions of the different fixed points.  In region C, only the all-cooperators fixed point $q^* =1$ is stable, while in region D, only the all-defectors fixed point $q^* =0$ is stable. In region C+D both fixed points are stable. The regions C and C+D are separated by the curve $r_s=r_s(c)$ and the regions D and C+D are separated  by the curve $r_a=r_a(c)$.
 }  
\label{fig:1}  
\end{figure}
%-----------------------------------------------------

The  results of the stability  analysis are conveniently summarized in a phase diagram in the plane $(c,r)$  as shown in Fig.\  \ref{fig:1} for $n=4$ and $n=8$. Only the fixed point $q^*=1$ is stable in  region C and only the fixed point $q^*=0$ is stable in  region D.  In  region C+D we have the bistability situation, where the outcome of the dynamics depends on the initial frequency of cooperators $q_0$. In particular, the unstable coexistence fixed point  $ 0< q^*<1$  determines the domains of attraction of the two stable fixed points: the dynamics leads to the all-cooperators regime when $q_0>q^*$ and to the all-defectors regime when $q_0<q^*$. An important feature of these phase diagrams, which also holds for general group sizes $n>2$, is that there is no stable coexistence solution. In fact, a stable coexistence  fixed point  would occur if $r_s < r_a$ in the limit $c \to 0$, so there would be a region where neither of the extreme fixed points is stable, but we have analytically ruled out this possibility by showing that $r_a \to r_s$ in this limit.

Another important limit where we can explicitly rewrite Eq. (\ref{Fp1g}) is for $c=1$, which corresponds to the worst-case scenario for cooperation and will be the main focus of our analysis henceforth.   In this case, we find
\begin{equation}
F_n'(1) = n - 1 + \frac{1}{2r}\frac{ n-1 - r \left ( n-2 \right )
- r^2 \left ( 2n^3 -10n^2 + 18n -11 \right )}{n-1 + r \left ( n^2 -3n + 3 \right )} ,
\end{equation}
so that the condition $|F_n'(1)| < 1$ is satisfied for $r > r_a$, where $r_a$ is the sole
positive solution of the quadratic equation
\begin{equation}\label{ra_quad}
r_a^2 - r_a \left ( 2n^2 -7n + 6 \right )   - \left (n - 1 \right ) = 0.
\end{equation}
For large $n$ we have $r_a \approx 2n^2$, and for the group sizes  used in drawing the phase diagrams in Fig. \ref{fig:1}  we have $r_a \approx 10.29$ for $n=4$ and $r_a \approx 78.09$ for $n=8$.

%-----------------------------------------------------
\begin{figure}[th] 
\centering
 \includegraphics[width=.8\columnwidth]{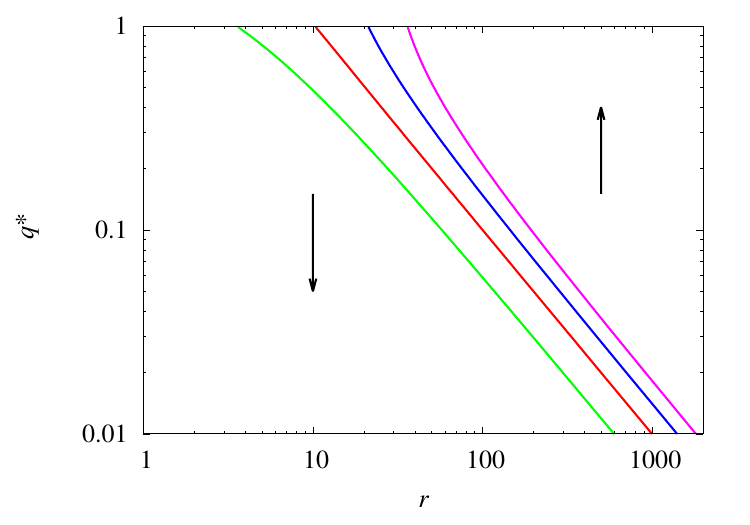}  
\caption{Unstable coexistence fixed point $q^*$  as a function of the amplification factor $r$ for $c=1$ and (from left to right) $n=3,4,5,6$. If $q_0> q^*$  the dynamics is driven to the all-cooperators fixed point and to the all-defectors fixed point if $q_0< q^*$, as indicated by the arrows.  The value of $r$ at which $q^*=1$ signals the instability of the all-cooperators fixed point and is given by the solution of Eq. (\ref{ra_quad}). The  all-defectors fixed point is  stable for these parameter settings.
 }  
\label{fig:2}  
\end{figure}
%-----------------------------------------------------

Figure \ref{fig:2} shows the unstable coexistence fixed point $q^*$ for $c=1$ and several group sizes $n$. In case of bistability, this fixed point determines the domains of attraction of the two stable fixed points. The coexistence and all-cooperators fixed points coincide  (i.e., $q^*=1$) at $r=r_a$, which is given for $c=1$  by the solution of Eq. (\ref{ra_quad}).  We use a log-log scale in this figure to emphasize the power-law decay of $q^*$ with increasing $r$. More explicitly, for $c=1$ and large $r$ we have an analytical expression for the unstable coexistence fixed point, viz. $q^* \approx
 (4n-6)/r $, which is in perfect agreement with the numerical results shown in Fig.\ \ref{fig:2}. This result is important because it shows that at maximum cost to cooperators (i.e., $c=1$), they can take over the population even at low initial frequencies, provided that the amplification factor $r$ is sufficiently large.

 \section{Simulations for finite number of groups}\label{sec:fin}
 
 Here we consider the robustness of our deterministic results to demographic noise, which arises when the
 population is finite, i.e., when $G$ is finite.   We focus only in  the case of biparental sexual reproduction, which can maintain a stable population of cooperators in the limit $G \to \infty$, as shown in the previous section. 
 
 %-----------------------------------------------------
\begin{figure}[t] 
\centering
 \includegraphics[width=1\columnwidth]{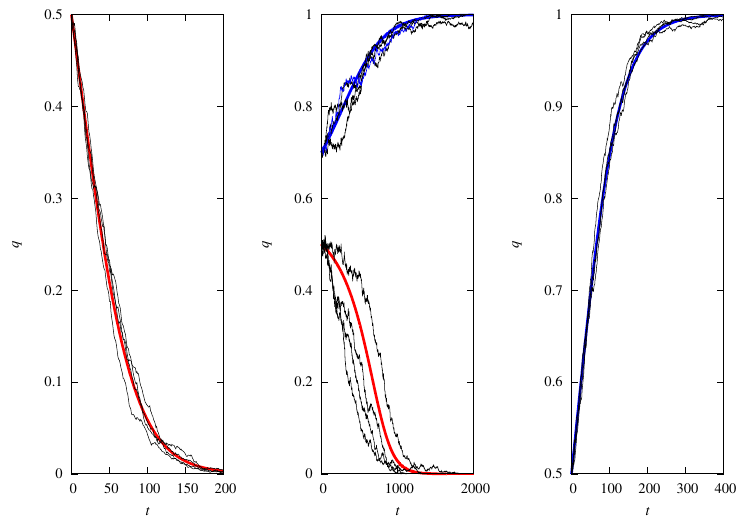}  
\caption{Frequency  of  cooperators as a function of time for  $r=5$ and $q_0=0.5$  (left panel),  $r=15$, $q_0=0.5$ and $q_0=0.7$ (middle panel),
$r=25$ and $q_0=0.5$  (right panel).
The jagged thin black curves are   runs of the simulations  for $G=10^4$ and the smooth thick colored curves are  the deterministic results.
 The other parameters are $n=4$ and $c=0.05$. 
 }  
\label{fig:3}  
\end{figure}
%-----------------------------------------------------
 
 To validate our analytical formulation, we compare in Fig.\ \ref{fig:3} the results obtained by iterating the recurrence (\ref{map}) and the simulation results for $G=10^4$ and $c=0.05$. We choose this particular small value of the contribution to the public goods because by varying $r$ we can go through the three different equilibrium regimes shown in the phase diagram of Fig.\ \ref{fig:1}. In addition to the good agreement between the analytical results and the individual runs of the simulations (the figure shows four runs for each parameter setting), we can see that the fluctuations due to demographic noise are higher in the bistability regime (middle panel of Fig.\ \ref{fig:3}), where the dynamics take a much longer time to reach equilibrium than in the other two regimes. Moreover, the  convergence to the all-defectors fixed point (left  panel of Fig.\ \ref{fig:3}) is  typically much faster than to the all-cooperators fixed point  (right panel of Fig.\ \ref{fig:3}).   For not too large $G$, in the bistability regime we can have fixation to different fixed points even starting from the same initial condition $q_0$ due to demographic noise. This is the problem we address next, considering a worst-case scenario for cooperation by setting $c=1$.

\begin{figure}[th]
\centering
\includegraphics[width=.8\columnwidth]{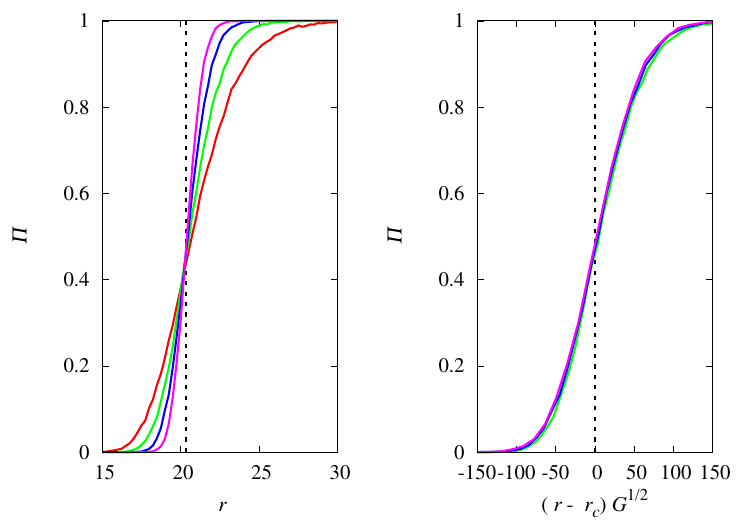}
\caption{Probability of fixation $\Pi$ of the cooperators as a function of the amplification factor $r$ (left panel)  for (from bottom to top at $r=25$) $G=400,800,1600, 3200$. The vertical dashed line indicates the threshold  $r_c = 20.37$  beyond which the all-cooperators fixed point attracts the orbits starting at $q_0= 0.5$ for $G \to \infty$. The right panel shows  $\Pi$ as a function of the scaled variable $(r-r_c)  G^{1/2}$ for $G= 800, 1600, 3200$.  The other parameters are $n=4$ and $c=1$. 
} 
\label{fig:4}
\end{figure}

To  quantify the effect of demographic noise, we consider the probability of fixation of the cooperators  $\Pi$, which is estimated as the fraction of $10^4$ independent runs for which this strategy fixates, i.e., for which  the stochastic dynamics is attracted to the all-cooperators absorbing state.  In the initial population, each individual is randomly assigned to one of the strategies with equal probability, corresponding to the choice $q_0 =0.5$ in the deterministic analysis.  In the bistability regime,  the relevant deterministic result is given in Fig.\ \ref{fig:2}, which can be interpreted in two different ways: for fixed $r$ there is a minimum initial frequency of cooperators  $q_0 =q^*$ above which the recurrence (\ref{map}) converges to the all-cooperators fixed point, and for fixed $q_0$ there is a minimum value of $r=r_c$  above which the recurrence (\ref{map}) converges to the all-cooperators fixed point. Here we choose the later interpretation and, in particular, for $q_0=0.5$ and $n=4$ we have  $r_c = 20.37$.  Accordingly, Fig.\  \ref{fig:4} shows  the probability of fixation of the cooperators as function of the amplification factor $r$. As noted above, for the same parameter setting, the demographic noise can tilt the dynamics into one absorbing state or the other.  The strength of this effect depends on the  number of groups and on the proximity of $r$ to $r_c$. In fact, the scaling assumption $\Pi \approx (r-r_c)G^{1/2}$ perfectly describes this dependence  for large $G$, implying that the width of the transition region around $r_c$ shrinks as $G^{-1/2}$ as $G$ increases (see \cite{Binder_1985,Campos_1999} for other applications of curve collapsing to characterize threshold phenomena).

%------------------------------------------------------------------------------------------------------------
%------------------------------------------------------------------------------------------------------------
%

\begin{figure}[ht]
\centering
\includegraphics[width=.8\columnwidth]{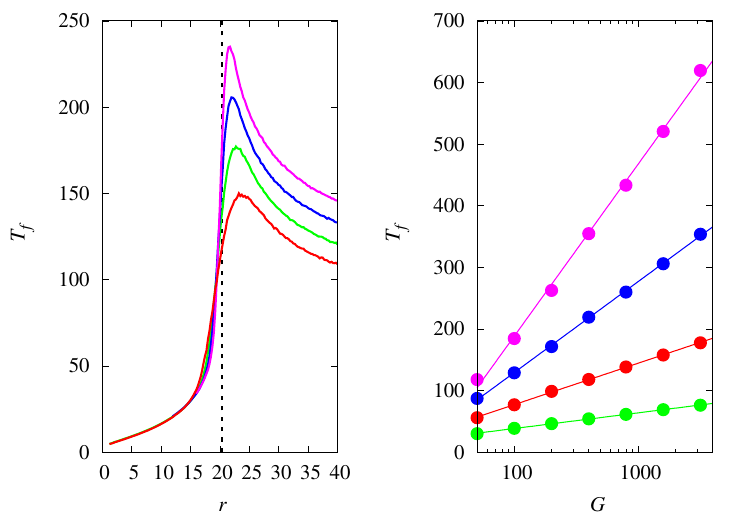}
\caption{Mean time for fixation of cooperators or defectors   $T_f$ as a function of amplification factor $r$ (left panel) for $n=4$ and (from bottom to top at $r=25$) $G=400,800,1600, 3200$. The vertical dashed line indicates the threshold  $r_c = 20.37$  beyond which the all-cooperators fixed point attracts the orbits starting at $q_0= 0.5$ for $G \to \infty$. The right panel shows $T_f$ at $r=r_c$  as function of $G$  for (from bottom to top) $n=3,4,5,6$. The lines are the fits  $T_f = b_f + a_f  \ln G$  where $b_f$ and $a_f$ are fit parameters.  
The  public goods contribution is  $c=1$.
} 
\label{fig:5}
\end{figure}

As already hinted at in Fig.\ \ref{fig:3}, another key quantity to characterize the stochastic dynamics is  the  mean  fixation time $T_f$, i.e., the mean time for the stochastic dynamics to reach the all-cooperators  or the all-defectors absorbing  states. Figure \ref{fig:5} shows that  in the all-cooperators regime ($r > r_c$) the fixation time diverges with $ a_g \ln G$, where the prefactor $a_g =a_g(n) $ is  an increasing function of the group size $n$. However, in the  all-defectors regime ($r<  r_c$) the fixation time tends to a finite limit as $G$ increases. Note that for finite $G$, the maximum of $T_f$ does not occur at $r = r_c$, but it moves in the direction of the threshold $r_c$  as $G$ increases.

%------------------------------------------------------------------------------------------------------------
%------------------------------------------------------------------------------------------------------------
%

\begin{figure}[ht]
\centering
\includegraphics[width=.8\columnwidth]{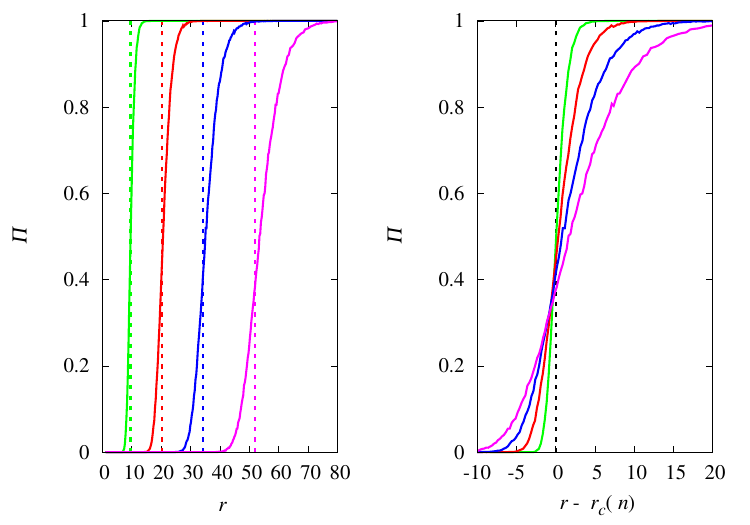}
\caption{Probability of fixation of the cooperators as a function of the amplification factor $r$ for (from left to right at $\Pi=0.8$) $n=3,4,5,6$ (left panel). The vertical dashed lines indicate the thresholds  $r_c(3) = 9.62$, $r_c(4) = 20.37$,  $r_c(5) = 34.32$ and $r_c(6) = 51.83$ beyond which the all-cooperators fixed point attracts the orbits starting at $q_0 = 0.5$  for $G \to \infty$. The right panel shows the same data with the curves shifted by $r_c(n)$.  The other parameters are $G=400$ and $c=1$.
} 
\label{fig:6}
\end{figure}

Figure \ref{fig:6} shows the effect of the group size $n$ on the threshold transition for finite $G$.   Although Fig.\ \ref{fig:2} already showed that the threshold $r_c$ increases with $n$, we can now see that the width of the transition region also increases with $n$, making the fixation of cooperators less likely for $r > r_c$ and finite $G$.  Indeed, the right panel of Fig.\ \ref{fig:5} shows that the mean fixation time also increases with $n$, confirming that increasing group size makes within-group competition more favorable for defectors, reducing the chances of the cooperators  to contribute to the offspring pool. Finally, we note that although in most of the results in this section we have fixed the initial frequency of cooperators at $q_0=0.5$, we get qualitatively similar results if we decrease $q_0$, provided that we increase the range of $r$ according to Fig.\ \ref{fig:2}.

%------------------------------------------------------------------------------------------------------------
%------------------------------------------------------------------------------------------------------------

\section{Discussion}\label{sec:disc}

There is a consensus that strong altruism, where  a cooperator does not benefit from her costly altruistic behavior,  cannot evolve in randomly formed, transient groups unless there is some sort of positive assortment among cooperators \cite{Hamilton_1975,Wilson_1975,Wilson_1990,Fletcher_2004,Nunney_1985,Maynard_1998}  or punishment for defectors \cite{Boyd_1992,Hannelore_2010,Sigmund_2010}.  Note that  reputation and reciprocity foster cooperation through the creation of positive assortment among players \cite{Wang_2023}.  Even with these fixes, one difficulty remains: the so-called `origination problem', i.e. how cooperation can take off from an initial low frequency of cooperators. 

A word on the terminology of weak and strong altruism is in order. Whether we are dealing with weak or strong altruism is determined by the fitness of the altruist or cooperator, Eq. (\ref{fa}): if the benefit she receives from her own contribution is greater than or equal to the cost of the altruistic act, then we have weak altruism, otherwise we have strong altruism. This terminology considers only first-order effects, the direct effect of the altruist on herself.   There are also second-order or indirect effects, where the altruist helps others who in turn help the altruist.  The various prescriptions for the emergence of cooperation in the strong altruism scenario (e.g., positive assortment, punishment, and reputation) involve playing with these second-order effects.  Of course, if such a prescription works, it is because the benefits of the second-order effects exceed the costs of the altruistic acts. Thus, including these indirect effects in the benefits accruing to the altruist leads to the conclusion that, by definition, cooperation cannot evolve in the strong altruism scenario.

The main purpose of this paper was to highlight the importance that the mode of reproduction can have on the fate of strong altruism in populations that are frequently divided into small temporary groups,  in which  individuals interact locally and fitness-related processes such as mating take place. The selection dynamics resulting from the different forms of reproduction considered here were solved analytically in the limit of infinitely many groups, assuming, as usual, that individuals colonize the  groups according to a binomial distribution, so that cooperators do not cluster into groups, and that local interactions follow the others-only variant of the $n$-prisoner's dilemma. Under biparental sexual reproduction, the results are quite striking, challenging conventional wisdom by showing that strong altruism can indeed evolve in the absence of assortment and punishment  in temporary groups if the  amplification factor $r$ is sufficiently large.   The basic  reason is that biparental sexual reproduction may enable a resident population of cooperators to resist invasion by rare defectors, since the fitness advantage of a single defector in a group with $n-1$ cooperators is smothered by the requirement that the recombinant offspring chromosome has a one-half chance of being a cooperator in any mating involving the defector.  However, and most surprisingly,  biparental sexual reproduction can also enable the invasion of a resident population of defectors by rare cooperators if the cost of cooperation $c$  is not large, thus solving the problem of origination. These results contrast strikingly with asexual reproduction and sexual reproduction with self-mating modes, for which cooperation is always doomed and the amplification factor $r$ can only slow the rate at which cooperators disappear from the population \cite{Hamilton_1975}. Although our conclusions apply strictly to populations of monoecious individuals, where each individual can produce both male and female gametes but they cannot self-fertilize, they also apply to dioecious populations, where individuals can produce either male or female gametes (see \ref{ref:A}). In particular, we find that dioecy promotes cooperation.

Our main findings are summarized in Fig.\ \ref{fig:1}, which shows a region, labeled C,  where cooperation is the only evolutionary outcome. For large group size $n$,  this region exists for  $c< 1/2n$ and $r > 2n^2$.  The only modification we have made to Hamilton's original model \cite{Hamilton_1975} is to change asexual reproduction to biparental sexual reproduction, which is how most sexual reproduction occurs \cite{Williams_1975}. This result provides a neat solution to both the problem of origination and the conditions under which strong altruism can evolve,  thus avoiding  the `inverse genetic fallacy' \cite{Field_2001}, i.e. the inappropriate attribution of mechanisms that might support cooperation  (e.g., positive assortment and punishment) to the explanation of its origin.  For large cooperation costs (i.e., $c \approx 1$) and large $n$, we find a wide region in the parameter space, labeled C+D in Fig.\ \ref{fig:1}, bounded at the bottom by $r=2n^2$, where cooperation can evolve provided that $q_0 > 4n/r$, where $q_0$ is the initial frequency of cooperators.  These estimates for large $n$ are worst-case scenarios for cooperation, since increasing group size $n$ favors defection.  In addition, we have shown that the results of our analytical study for infinitely many groups are robust to the demographic noise that arises when the number of groups is finite.

How realistic are the conditions in our model for the evolution of cooperation in nature? One possible example that comes to mind is viral co-infection, which complicates the symptoms and diagnosis of disease \cite{Du_2022}, and also leads to some interesting evolutionary outcomes \cite{Turner_1999,Turner_2003,Elena_2014}.  Viral co-infections can be modeled by analogy with Wilson's  trait groups \cite{Szathmary_1992}. In RNA viruses, recombination can occur via a copy-choice mechanism during viral replication, in which the viral RNA polymerase switches templates during strand synthesis \cite{Jarvis_1991}. 
Virus-virus interactions after co-infection can be modeled using game theory \cite{Elena_2014}, as in the classic example of the RNA phage $\phi 6$, where competitive interactions were shown to follow the rules of the prisoner's dilemma \cite{Turner_1999}. In this case, the fixation of high-multiplicity phages led to a decrease in population fitness. To escape the dilemma, it has been suggested that fitness payoffs evolve from the prisoner's dilemma to the snowdrift game, where coexistence between cooperators and defectors is possible \cite{Chao_2017}.  How general this result can be in viral co-infections is unknown, but as shown here, there may be circumstances where the payoffs are consistent with the prisoner's dilemma and yet cooperation can still evolve.

A noticeable feature of our study is the high values of the amplification factor $r$, which measures the benefit of the cooperators to the group,  necessary for the emergence of cooperation in the scenario of biparental sexual reproduction. This benefit is  the mean number of offspring produced by the  group, which grows linearly with $r$ according to Eqs. (\ref{pi}) and (\ref{av_of}).   The population biology of early replicators, in particular the enzyme production problem \cite{Michod_1983,Mariano_2024}, provides a scenario where $r$ is expected to be very large.  Here the cooperators are enzyme producers and the defectors benefit from the enzyme but do not encode it (so the defectors are shorter than the cooperators and therefore replicate faster).  The fitness used to model this problem is exactly the weak altruism version of the n-person prisoner's dilemma, and since enzyme-catalyzed reactions are often many orders of magnitude faster than their non-catalyzed counterparts, it follows that $r$ must be very large \cite{Michod_1983}.  In fact, we note that the interaction between enzyme and replicator is a hypercyclic coupled biosynthetic process, and the classical experiments on phage-infected bacteria show that during the early stage of viral infection there is a hyperbolic increase in the viral RNA synthesis rate, which is a clear sign of hypercyclic coupling \cite{Eigen_1981,Gebinoga_1997}. This suggests that the role of hypercyclic coupling in early molecular evolution is akin to its role in viral evolution today, which, as discussed above, provides a biological  context for our model.

Finally, it is an open question whether our conclusions apply to diploid sexual organisms under the more realistic assumption that altruistic and selfish strategies are quantitative traits rather than discrete behaviors. 
Indeed, it has been argued that if strategies exist as quantitative traits, rather than as discrete traits  as ours and most models assume, then   there are conditions under which defectors can be invaded by  rare cooperators, at least in a weak altruism scenario \cite{Wilson_1997}. 
  As Maynard Smith wrote \cite{Maynard_1982}: ``Most populations of interest have sexual diploid inheritance. In many cases this does not matter.''  Apart from the obvious case where a pure phenotypic strategy is produced by a heterozygous genotype and cannot be fixed in a random mating population because of Mendelian inheritance, we believe there is no simple answer to the question of whether it does or does not matter.

\section*{Acknowledgments}
JFF is partially supported by  Conselho Nacional de Desenvolvimento Ci\-en\-t\'{\i}\-fi\-co e Tecnol\'ogico (Brazil), grant number 305620/2021-5. MS is funded by  grant PID2021-127107NB-I00 from Ministerio de Ciencia e Innovaci\'on (Spain) and  grant 2021 SGR 00526 from Generalitat de Catalunya.

\section*{Declaration of interest:} None.

\appendix

\renewcommand{\theequation}{A.\arabic{equation}}
\setcounter{equation}{0}
\setcounter{figure}{0}

\section{}\label{ref:A}

In the main text we considered a population composed of monoecious individuals, i.e. individuals that have both male and female reproductive organs, and can produce gametes of both sexes.   Here we consider a finite  population of dioecious individuals, so that only males and females can mate. We also assume equal sex ratio and no correlation between sex and strategies.  The assumption of equal sex ratios applies to the population as a whole, not to the sex ratio in a particular group, which may vary according to binomial sampling. The model is considerably more complicated than the monoecious case discussed in the main text, so we limit our analysis to simulations of finite populations, which is sufficient to show that our main conclusions also hold in dioecious populations. Surprisingly, we find that cooperation is more likely to evolve in a population of dioecious individuals than in a population of monoecious individuals.

First, all $G \times n$ individuals are randomly assigned a strategy (cooperator or defector) and a sex (male or female). Then, $G$ groups of size $n$ are formed by randomly selecting individuals without replacement.  Fitness do not depend on the sex of the individuals so
the fitness of a cooperator in a group with $i-1$ cooperators $f_c(i)$ is given by Eq.  (\ref{fa}), the fitness of a defector in a group with $i$ cooperators $f_d(i)$ is given by Eq. (\ref{fs}), and the mean fitness of a group $\pi(i)$ with $i$ cooperators is given by Eq. (\ref{pi}). The number of offspring $m$ produced by a group with $i$ cooperators is a Poisson random variable with mean $\pi(i)$ (see Eq. \ref{Poisson}) if the group contains individuals of both sexes, otherwise $m=0$.  This is the first major difference from the monoecious case, discussed in the main text. The second major difference, of course, is the choice of parents, who must now be of different sexes.

The strategy of  an offspring  is determined by the strategies of their parents.  First, we choose a father from among the males in the group with a probability given by his relative fitness. A male's relative fitness is given by the ratio of his fitness to the total fitness of the males in the group, so the choice of a father involves only competition between males.  Then we choose a mother from among the females in the group with probability proportional to her fitness, following the same procedure described above: the choice of a mother involves a competition between females only. The offspring will inherit either the strategy of the father or the strategy of the mother with equal probability.  To avoid correlations between sex and strategy, the sex of the offspring is randomly determined with equal probability.

%------------------------------------------------------------------------------------------------------------
\begin{figure}[ht]
\centering
\includegraphics[width=.8\columnwidth]{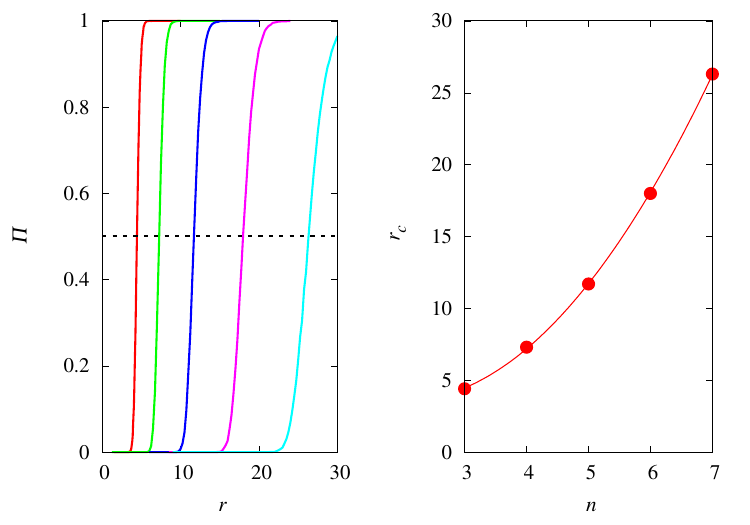}
\caption{Probability of fixation of the cooperators as a function of the amplification factor $r$ for (from left to right) $n=3,4,5,6, 7$ (left panel). The right panel shows the  rough estimate of the threshold $r_c$ obtained with  the condition $\Pi=0.5$.  The line is the fit $r_c = 0.91 - 3.63 n + 7.20 n^2$. The other parameters are $G=400$ and $c=1$.
} 
\label{fig:A1}
\end{figure}
%------------------------------------------------------------------------------------------------------------

The offspring pool is formed by the offspring produced by each of the $G$ groups, and the new groups are formed by sampling with replacement $G \times n$ individuals from this pool, as done for the population of monoecious individuals. The procedure is then repeated until one of the strategies becomes fixed. The left panel of Fig. \ref{fig:A1} shows the probability of fixation of the cooperators for different group sizes $n$, which should be compared with the left panel of Fig. \ref{fig:6}. The remarkable result is that the transition to the all-cooperators regime occurs for a smaller value of the amplification factor $r$ compared to the results for the monoecious individuals.  A rough estimate of the threshold $r_c$, obtained by finding the value of $r$ at which $\Pi=0.5$, is shown in the right panel of Fig. \ref{fig:A1}. The results show that $r_c$ increases with $n^2$, similar to the case of  biparental sexual reproduction  discussed in the main text. We have also found that the all-cooperators regime is the only absorbing state for $n=2$, as expected.  In fact, as argued before, the only advantage of a defector is to share the group with a cooperator, but this leverage is lost because the offspring have a 50\% chance of inheriting the cooperative trait.

Why is the threshold $r_c$ lower for dioecious individuals? To gain some insight into this, we recall that increasing group size hinders cooperation. Since we restrict intragroup competition only between males and between females, the effective group size perceived by each group member is reduced, which explains our results.

%\section*{References}

\end{document}